\documentclass[aps,showpacs,preprintnumbers,nofootinbib,amsmath,amssymb]{revtex4-1}
\usepackage{slashed,color,amsmath,amssymb,mathrsfs}
\usepackage{graphicx}
\usepackage{ulem}
\usepackage{hyperref}
\usepackage{longtable}
\usepackage{array}

\allowdisplaybreaks
\begin{document}
\preprint{}
\title{Derivation of the effective Chiral Lagrangian for pseudoscalar,\\ scalar, vector and axial-vector  mesons from QCD}

\author{Ke Ren}
\email[]{kren3@163.com}
\affiliation{Department of Physics, Tsinghua University, Beijing 100084, P.R.China}
\author{Hui-Feng Fu}
\email[]{huifengfu@jlu.edu.cn}
\affiliation{Center for Theoretical Physics, College of Physics, Jilin University, Changchun 130012, P.R.China}

\author{Qing Wang}
\email[Corresponding author:~]{wangq@mail.tsinghua.edu.cn}
\affiliation{Department of Physics, Tsinghua University, Beijing 100084, P.R.China}
\affiliation{Center for High Energy Physics, Tsinghua University, Beijing 100084, P.R.China}
\affiliation{Collaborative Innovation Center of Quantum Matter, Beijing 100084, P.R.China}

\begin{abstract}
A previous formal derivation of the effective chiral Lagrangian for low-lying pseudoscalar mesons from first-principles QCD without approximations [Wang et al., Phys. Rev. D61, (2000) 54011] is generalized to further include scalar, vector, and axial-vector mesons. In the large $N_c$ limit and with an Abelian approximation, we show that the properties of the newly added mesons in our formalism are determined by the corresponding underlying fundamental homogeneous Bethe--Salpeter equation in the ladder approximation, which yields the equations of motion for the scalar, vector, and axial-vector meson fields at the level of an effective chiral Lagrangian. The masses appearing in the equations of motion of the meson fields are those determined by the corresponding Bethe--Salpeter equation.
 \end{abstract}
 \maketitle
\section{Introduction}

Effective chiral Lagrangian theory is well known to be a very useful formalism to describe low-lying mesons, which enables us to avoid difficulties in nonperturbative QCD in the low-energy regime. We often dismiss the importance of derivations of the effective chiral Lagrangian from first-principles QCD theory and treat this derivation as a purely theoretical exercise. This, of course, is partly due to the difficulties in the derivation, but the more important reason may be because the effective chiral Lagrangian has the same chiral and discrete symmetries as the original QCD, regardless whether the Lagrangian is derived or simply written down. The same form of the effective chiral Lagrangian is always obtained, and in this sense, the derivation seems to lose its value, because in terms of symmetry constraints, it suffices to write all necessary terms of the effective chiral Lagrangian. Then, why do we need to pay a high price and play with this very difficult derivation? There are three reasons. First, the derivation offers us a systematic definition of low-energy constants (LECs) in terms of the underlying QCD, which conventionally can only be done for specific LECs by choosing some special processes or physical quantities and demanding a match of the computed results from the effective chiral Lagrangian and from QCD. Based on this systematic and fundamental expression of LECs at the QCD level, one can develop some first-principles QCD computation techniques for LECs. This provides us with a possible solution to the LEC number problem encountered in effective chiral Lagrangian theory, for which if we go to high order in the low-energy expansion, the number of LECs increases very rapidly. It is impossible to fix all of them purely from experiment. Second, the derivation often provides us with a compact expression for the effective chiral Lagrangian, which corresponds to re-summing the infinite series of the low-energy expansion. This offers a way to go beyond the traditional low-energy expansion for the effective chiral Lagrangian and enlarge the available scope of our calculation. Third, the detailed procedure of the derivation in general is independent of various symmetry arguments which are relied heavily by traditional effective chiral Lagrangian. Often this reveals some unexpected correlations between different objects, which will make us go beyond conventional pure symmetry discussions and deepen our understanding of the structure of the QCD. For our derived effective chiral Lagrangian, although the Lagrangian itself is phenomenological, the detailed structure of the final expression includes all necessary information of the underlying QCD. Although the derivation of the effective chiral Lagrangian from QCD is of such importance, unfortunately, it often is very difficult involving a lot of mathematical and nonperturbative techniques to treat. The naive belief is that, if one can successfully handle these difficulties, compare to derive the effective chiral Lagrangian, directly using these techniques to compute low-energy physics quantities and processes may be more efficient. For these reasons, very little work on a reliable derivation has appeared in the literature.

Seventeen years ago, in Ref.~\cite{WQ0}, we built a formalism to derive the chiral Lagrangian for low-lying pseudoscalar mesons from first-principles QCD without taking approximations. The derivation is based on the standard generating functional of QCD with external bilinear light-quark fields sources in the path integral formalism. The integrations of gluon and heavy-quark field are performed formally by expressing the result in terms of the physical Green functions of the gluon. To integrate over the left light-quark fields, we introduce a bilocal auxiliary field $\Phi(x,y)$ representing the light mesons. We then developed a consistent way of extracting the local pseudoscalar degree of freedom $U(x)$ in $\Phi(x,y)$ integrating out the other degrees of freedom. With certain techniques, we worked out the explicit $U(x)$ dependence of the effective action up to the $p^4$ terms in the momentum expansion, which leads to the desired chiral Lagrangian in which all LECs of the theory are expressed in terms of Green's functions for certain quarks in QCD. The final result can be regarded as the fundamental QCD definition of the LECs. With some approximation, we found our QCD definition of the $p^2$-order coefficient $F_0^2$ recovers just the well-known approximate result given by Pagels and Stokar\cite{PS}.

Further developments of this work moved along two directions. One was in terms of compact result obtained in Ref.~\cite{WQ0} from QCD. We continued to develop various techniques to calculate LECs of the pseudoscalar meson chiral Lagrangian that included LECs of order $p^4$ \cite{WQ02PRD}, and various versions of order $p^6$ \cite{WQ10normal,WQ10anomalous,WQ13tensor,WQ15update}. To date though, we still use an ansatz to represent the real solution of the external source-dependent Schwinger--Dyson equation to simplify the computations. Because of its rough approximation, it is criticized by becoming phenomenological model\cite{WQ02PLB} and not derived from a first-principles calculations. We are nevertheless improving our techniques step by step toward a final realization of a true first-principles computation. The other direction developed extends the content of the formalism from pure pseudoscalar mesons to include other meson species, in particular, vector mesons in Ref.~\cite{WQ00vector}, and the $\eta'$ meson in Ref.~\cite{WQ00etap}. In the following year, we changed from QCD to a general technicolor model deriving the effective chiral Lagrangian for its pseudo-Goldstone bosons in Ref.~\cite{WQ01tech}. This direction of development was stopped in 2001,  because we have found a key obstacle to a result given in \cite{WQ00vector}. There is inevitably a free dimensional parameter $\mu$ appearing in our derivation, and it seems that the masses of the vector bosons depending on this $\mu$ parameter can take arbitrary values . This contradicts our expectation because we know that QCD determines the property of vector mesons non-perturbatively via the corresponding Bethe--Salpeter equation (BSE)\cite{BSvector}, which will fix the vector meson mass without any arbitrariness. For a correct formalism, we should automatically produce this BSE, which fixes the amplitude of vector boson. The situation for vector mesons contrasts that of pseudoscalar mesons, where spontaneous chiral symmetry breaking at the chiral limit sets their masses to zero because of their Goldstone nature. As long as our formalism retains the correct behavior under  spontaneous chiral symmetry breaking, the masses of pseudoscalar mesons are automatically zero. Beyond pseudoscalar mesons, we do not have this Goldstone-theorem argument and the masses of mesons no longer vanish even at the chiral limit. Hence, producing the correct value of these nonzero meson masses becomes a real challenge. It is a sign that our derivation goes beyond the constraint of chiral symmetry and its breaking. Here we stress the importance of producing the correct meson masses or more generally the LECs of the effective chiral Lagrangian in our derivation. Without this we cannot trust the validity of the formalism we have built up for the derivation. In fact, in the mid-1980s, there were many articles claiming that they have achieved the derivation in terms of the non-minimal anomaly of QCD \cite{anomaly1,anomaly2,anomaly3,anomaly4,anomaly5,anomaly6,anomaly7}. The general criticisms of the work are that the derivations do not involve QCD dynamics because when the quark--gluon interaction is switched off, nonzero LECs remained. As a result, the corresponding derivations could not produce the correct $F_0^2$ (most of the attempts just gave zero) and the $p^4$-order LECs $L_7$ and $L_8$ gave the wrong signs. In comparison, our work \cite{WQ0} did give the correct $F_0^2$ predicted by Pagels and Stokar\cite{PS} and produced the correct $L_7$ and $L_8$ signs, which boosted confidence in our formalism and derivations. The problem became the evaluation of the correct value of the meson mass or its other properties to judge the reliability of the formalism and derivation, before solving the problem of arbitrary mass of the vector meson. Because of the existence of the free parameter $\mu$, we lacked assurance to develop the formalism further for mesons of heavier mass and had no idea how to avoid the effect of this arbitrary parameter on vector meson masses or how to generate the BSE within our formalism for vector mesons. Indeed, we doubted whether our formalism for the vector meson was correct. Considering that the vector meson is of the lowest mass above those of the three pseudoscalar flavor mesons, we did not expect our formalism to work for other more heavier mesons if it did not work for vector mesons. The purpose of this work is to solve this problem and show that the formalism set up in Ref.~\cite{WQ00vector} is correct. The free parameter $\mu$ does not have effect on vector meson masses and we can reproduce the standard BSE for vector mesons \cite{BSvector} in our formalism as well as the correct physical mass in the effective chiral Lagrangian. Our discussion can be further generalized to scalar and axial-vector mesons.

This paper is organized as follows, in Sec.~II, we review the previous work for pseudoscalar mesons \cite{WQ0} and vector mesons \cite{WQ00vector}; in Sec.~III, in the large $N_c$ limit, which was not discussed previously, we show in the Abelian approximation that the formalism established can reproduce the correct ladder approximation BSE for vector mesons and the property of vector meson is independent of free parameter $\mu$. We find that in our formalism, BSE leads to the equation of motion (EOM) for vector mesons and the mass appearing in EOM is just that determined by BSE. In Sec.~IV, we generalize our formalism to further include scalar and axial-vector mesons. Sec.~V provides a summary and discussion.
\section{Review of previous derivations of the effective chiral Lagrangian for pseudoscalar and vector mesons}

In this section, we review the necessary parts of Refs.~\cite{WQ0,WQ00vector} for our later discussion in this paper. Consider a QCD-type gauge theory with $SU(N_c)$ local gauge symmetry and $N_f$ flavor quarks. We start from generating functional given in Ref.~\cite{WQ0},
\begin{eqnarray}
Z[J]&=&\int{\cal D}\psi{\cal D}\bar{\psi}\exp\bigg\{i\int d^4x\bar{\psi}(i\slashed{\partial}+J)\psi\bigg\}\int{\cal D}q{\cal D}\bar{q}{\cal D}A_\mu\Delta_F(A_\mu)\nonumber\\
&&\times\exp\bigg\{i\int d^4x\bigg[{\cal L}_{\mathrm{QCD}}(A)-\frac{1}{2\xi}[F^i(A_\mu)]^2-g{\cal I}_i^{\mu}A^i_\mu+\bar{q}(i\slashed{\partial}-M-g\slashed{A})q\bigg]\bigg\}.\label{Zdef}
\end{eqnarray}
The conventions and definitions for fields in Eq.~(\ref{Zdef}) are the same as those in Ref.~\cite{WQ0}. As done in Ref.~\cite{WQ0}, we integrate out gluon and quark fields, and integrate in a bilocal auxiliary field $\Phi^{\sigma\rho}(x,y)\sim(1/N_c)\bar{\psi}^{\sigma}_{\alpha}(x)\psi^{\rho}_{\alpha}(y)$, which represent the bilinear light-quark pair. The result is
\begin{eqnarray}
Z[J]&=&\int{\cal D}\Phi{\cal D}\Pi~e^{i\Gamma_1[J,\Phi,\Pi]}\;,\label{ZGamma1}\\
\Gamma_1[J,\Phi,\Pi]&=&N_c\bigg\{-i\mathrm{Tr}\ln[i\slashed{\partial}+J-\Pi]+\int d^4xd^4x'\Phi^{\sigma\rho}(x,x')\Pi^{\sigma\rho}(x,x')
+{\displaystyle\sum_{n=2}^{\infty}}\int d^4x_1\cdots d^4x_nd^4x'_1\cdots d^4x'_n\frac{(-i)^n(N_cg^2)^{n-1}}{n!}\nonumber\\
&&\times\overline{G}^{\sigma_1\cdots\sigma_n}_{\rho_1\cdots\rho_n}(x_1,x'_1,\cdots,x_n,x'_n)\Phi^{\sigma_1\rho_1}(x_1,x'_1)\cdots\Phi^{\sigma_n\rho_n}(x_n,x'_n)\bigg\}\;.\label{Gamma1def}
\end{eqnarray}
where $\overline{G}^{\sigma_1\cdots\sigma_n}_{\rho_1\cdots\rho_n}(x_1,x'_1,\cdots,x_n,x'_n)$ is a generalized Green function for gluon including in contribution from gluons and heavy quarks. Then introducing unitary Goldstone fields $\xi_R$ and $\xi_L$ by inserting in Eq.~(\ref{ZGamma1}) a series Goldstone-field integration as in Ref.~\cite{WQ00vector}, Eq.~(\ref{ZGamma1}) is changed to
\begin{eqnarray}
Z[J]&=&\int{\cal D}\Phi{\cal D}\Pi{\cal D}\xi_R{\cal D}\xi_L{\cal D}\Xi~\delta(\xi_R^\dag\xi_R-1)\delta(\xi_L^\dag\xi_L-1)\delta(\mathrm{det}\xi_R-\mathrm{det}\xi_L)\exp\bigg\{i\Gamma_1[J,\Phi,\Pi]\nonumber\\
&&+i\Gamma_I[\Phi]+iN_c\int d^4x~\mathrm{tr}_f\{\Xi[e^{-i\vartheta/N_f}\xi_R\mathrm{tr}_l(P_R\Phi^T)\xi_L^\dag-e^{i\vartheta/N_f}\xi_L\mathrm{tr}_l(P_L\Phi^T)\xi_R^\dag]\}\bigg\}\;,
\label{ZxiRL}
\end{eqnarray}
where
\begin{eqnarray}
e^{-i\Gamma_I[\Phi]}\equiv\prod_x\bigg[\{\mathrm{det}[\mathrm{tr}_lP_R\Phi^T(x,x)]\mathrm{det}[\mathrm{tr}_lP_L\Phi^T(x,x)]\}^{1/2}\int{\cal D\sigma}~\delta[(\mathrm{tr}_lP_R\Phi^T)(\mathrm{tr}_lP_L\Phi^T)-\sigma^\dag\sigma]\delta(\sigma-\sigma^\dag)\bigg]
\end{eqnarray}
is a compensation term to cancel the residual $\Phi$ field dependence due to Goldstone fields $\xi_R$ and $\xi_L$ integration. A further integration in the vector meson field $V_{\mu}^{ab}(x)$ by functionally inserting a constant integration in Eq.~(\ref{ZxiRL}) yields
\begin{eqnarray}
\int{\cal D}V~\delta\bigg[V_{\mu}^{ab}(x)+\frac{1}{4\mu^4}\{[\xi_L(x)P_R+\xi_R(x)P_L]\Pi(x,x)[\xi_R^\dag(x)P_R+\xi_L^\dag(x)P_L]\}^{(a\xi)(b\zeta)}(\gamma_\mu)^{\zeta\xi}\bigg]\;.
\end{eqnarray}
Note here a mass dimension parameter $\mu$ is inevitable. This is because the $\Pi$ field has mass dimension 5 and $V_\mu$ has dimension 1. To compensate the dimensional difference between $\Pi$ and $V_\mu$ fields, we have to match this with a constant of mass dimension 4, $\mu^4$. This parameter is apparently arbitrary. Nevertheless, the physical results of our formalism must be independent of the value of this parameter. We can treat this $\mu$ independence of the result as a check of the reliability of our formalism. Now we exponentiate the delta function by introducing a functional integration over another field $\tilde{V}^{\mu,ba}(x)$ and formally integrate out the fields $\Pi$ and $\Phi$, $Z[J]$, obtaining
\begin{eqnarray}
Z[J]=\int{\cal D}\xi_R{\cal D}\xi_L{\cal D}V{\cal D}\Xi{\cal D}\tilde{V}~\delta(\xi_R^\dag\xi_R-1)\delta(\xi_L^\dag\xi_L-1)\delta(\mathrm{det}\xi_R-\mathrm{det}\xi_L)
e^{i\Gamma_2[\xi_R,\xi_L,V,J,\Xi,\tilde{V},\Phi_c,\Pi_c]}\;,\label{ZV}
\end{eqnarray}
where
\begin{eqnarray}
e^{i\Gamma_2[\xi_R,\xi_L,V,J,\Xi,\tilde{V},\Phi_c,\Pi_c]}&=&\int{\cal D}\Phi{\cal D}\Pi~\exp\bigg\{i\Gamma_1[J,\Phi,\Pi]+i\Gamma_I[\Phi]\nonumber\\
&&+iN_c\int d^4x\mathrm{tr}_f\bigg[\Xi[e^{-i\vartheta/N_f}\xi_R\mathrm{tr}_l(P_R\Phi^T)\xi_L^\dag-e^{i\vartheta/N_f}\xi_L\mathrm{tr}_l(P_L\Phi^T)\xi_R^\dag]\bigg]\nonumber\\
&&+iN_c\int d^4x\bigg[
V_{\mu}^{ab}+\frac{1}{4\mu^4}[(\xi_LP_R+\xi_RP_L)\Pi(\xi_R^\dag P_R+\xi_L^\dag P_L)]^{(a\xi)(b\zeta)}(\gamma_{\mu})^{\zeta\xi}\bigg]\tilde{V}^{\mu,ba}\bigg\}\;.
\label{Gamma2def0}
\end{eqnarray}
$\Pi_c$ and $\Phi_c$ fields are
\begin{eqnarray}
\Pi_c=\frac{\int{\cal D}\Phi{\cal D}\Pi~\Pi~e^{\cdots}}{\int{\cal D}\Phi{\cal D}\Pi~e^{\cdots}},\hspace*{2cm}\Phi_c=\frac{\int{\cal D}\Phi{\cal D}\Phi~\Pi~e^{\cdots}}{\int{\cal D}\Phi{\cal D}\Pi~e^{\cdots}},\label{PicPhic0}\;,
\end{eqnarray}
where $\cdots$ is the exponent of the exponential in Eq.~(\ref{Gamma2def0}). $\Pi_c$ and $\Phi_c$ satisfy the following equations
\begin{eqnarray}
\frac{\partial\Gamma_2[\xi_R,\xi_L,V,J,\Xi,\tilde{V},\Phi_c,\Pi_c]}{\partial\Pi_c}=0,\hspace*{2cm}\frac{\partial\Gamma_2[\xi_R,\xi_L,V,J,\Xi,\tilde{V},\Phi_c,\Pi_c]}{\partial\Phi_c}=0\;.
\label{PicPhic-1-0}
\end{eqnarray}
We formally finish the integration over fields $\tilde{V}$ and $\Xi$ in Eq.~(\ref{ZV}),
\begin{eqnarray}
Z[J]= \int{\cal D}\xi_R{\cal D}\xi_L{\cal D}V~\delta(\xi_R^\dag\xi_R-1)\delta(\xi_L^\dag\xi_L-1)\delta(\mathrm{det}\xi_R-\mathrm{det}\xi_L)
e^{iS_{\mathrm{eff}}[\xi_R,\xi_L,V,J,\Xi_c,\tilde{V}_c,\Phi_c,\Pi_c]}\;,\label{ZV-1}
\end{eqnarray}
where
\begin{eqnarray}
e^{iS_{\mathrm{eff}}[\xi_R,\xi_L,V,J,\Xi_c,\tilde{V}_c,\Phi_c,\Pi_c]}
=\int{\cal D}\Xi{\cal D}\tilde{V}~
e^{i\Gamma_2[\xi_R,\xi_L,V,J,\Xi,\tilde{V},\Phi_c,\Pi_c]}\;.\label{Seffdef1}
\end{eqnarray}
The $\Xi_c$ and $\tilde{V}_c$ fields are
\begin{eqnarray}
\Xi_c=\frac{\int{\cal D}\Xi{\cal D}\tilde{V}~\Xi~e^{\cdots}}{\int{\cal D}\Xi{\cal D}\tilde{V}~e^{\cdots}},\hspace*{2cm}\tilde{V}_c=\frac{\int{\cal D}\Xi{\cal D}\tilde{V}~\tilde{V}~e^{\cdots}}{\int{\cal D}\Xi{\cal D}\tilde{V}~e^{\cdots}},\label{XictildeVc0}
\end{eqnarray}
where $\cdots$ is the exponent of the exponential in Eq.~(\ref{Seffdef1}). $\Xi_c$ and $\tilde{V}_c$ satisfy the following equations
\begin{eqnarray}
\frac{\partial S_{\mathrm{eff}}[\xi_R,\xi_L,V,J,\Xi_c,\tilde{V}_c,\Phi_c,\Pi_c]}{\partial\Xi_c}=0,
\hspace*{2cm}
\frac{\partial S_{\mathrm{eff}}[\xi_R,\xi_L,V,J,\Xi_c,\tilde{V}_c,\Phi_c,\Pi_c]}{\partial\tilde{V}_c}
=0\;.\label{XictildeVc-1-0}
\end{eqnarray}
With the help of Eqs.~(\ref{Seffdef1}), Eqs.~(\ref{XictildeVc-1-0}), (\ref{PicPhic-1-0}), (\ref{Gamma2def0}), and (\ref{Gamma1def}), we obtain,
\begin{eqnarray}
&&\hspace*{-1cm}\frac{\partial S_{\mathrm{eff}}[\xi_R,\xi_L,V,J,\Xi_c,\tilde{V}_c,\Phi_c,\Pi_c]}{\partial J^{\sigma\rho}(x)}\bigg|_{\xi_R,\xi_L,V~fix}\hspace*{-0.8cm}=
N_c\bigg[\overline{\Phi_c^{\sigma\rho}(x,x)+\frac{\delta(0)}{4\mu^4}\{[\xi_R^\dag(x)P_R+\xi_L^\dag(x)P_L]\slashed{\tilde{V}}(x)[\xi_L(x)P_R+\xi_R(x)P_L]\}^{\sigma\rho}}\bigg]\;,\label{SeffderiveJ}\\
&&\hspace*{-1cm}\frac{\partial S_{\mathrm{eff}}[\xi_R,\xi_L,V,J,\Xi_c,\tilde{V}_c,\Phi_c,\Pi_c]}{\partial V_{\mu}^{ab}(x)}\bigg|_{\xi_R,\xi_L,J~fix}=
N_c\overline{\tilde{V}^{\mu,ba}(x)}\;.\label{EMO-V-0}
\end{eqnarray}
Here we use a bar to denote the functional average over fields $\Xi$ and $\tilde{V}$,
\begin{eqnarray}
\overline{O(x)}\equiv\frac{\int{\cal D}\Xi{\cal D}\tilde{V}~O(x)~e^{i\Gamma_2[\xi_R,\xi_L,V,J,\Xi,\tilde{V},\Phi_c,\Pi_c]}}{\int{\cal D}\Xi{\cal D}\tilde{V}~e^{i\Gamma_2[\xi_R,\xi_L,V,J,\Xi,\tilde{V},\Phi_c,\Pi_c]}}\;,
\end{eqnarray}
where the appearance of the second term on the r.h.s of Eq.~(\ref{SeffderiveJ}) is proportional to $\delta(0)$ because the last term of r.h.s. of Eq.~(\ref{Gamma2def0}) contributes an extra linear $\Pi$ term to the exponential of the integrand of the path integral, which because of the integration over the $\Pi$ field changes the original constraint $\Phi^{\sigma\rho}(x,y)\sim(1/N_c)\bar{\psi}^{\sigma}_{\alpha}(x)\psi^{\rho}_{\alpha}(y)$ to
$\Phi^{\sigma\rho}(x,y)+\frac{\delta(x-y)}{4\mu^4}\{[\xi_R^\dag(x)P_R+\xi_L^\dag(x)P_L]\slashed{\tilde{V}}(x)[\xi_L(x)P_R+\xi_R(x)P_L]\}^{\sigma\rho}\sim(1/N_c)\bar{\psi}^{\sigma}_{\alpha}(x)\psi^{\rho}_{\alpha}(y)$; the derivation of $J^{\sigma\rho}(x)$ for the path integral is simply $\bar{\psi}^{\sigma}_{\alpha}(x)\psi^{\rho}_{\alpha}(x)$. Defining the rotated source and fields by
\begin{eqnarray}
&&J_{\Omega}(x)=[\xi_L(x)P_R+\xi_R(x)P_L][J(x)+i\slashed{\partial}][\xi_R^\dag(x)P_R+\xi_L^\dag(x)P_L],\nonumber\\
&&\Phi^T_{\Omega}(x,y)=[\xi_R(x)P_R+\xi_L(x)P_L]\Phi^T(x,y)[\xi_L^\dag(y)P_R+\xi_R^\dag(y)P_L],\nonumber\\
&&\Pi_{\Omega}(x,y)=[\xi_L(x)P_R+\xi_R(x)P_L]\Pi(x,y)[\xi_R^\dag(y)P_R+\xi_L^\dag(y)P_L],\label{rotation}
\end{eqnarray}
the fields $\vartheta$, $\Xi$, $V$, $\tilde{V}$ remain unchanged. We distinguish the fields after transformation using a subscript $\Omega$; we then have
\begin{eqnarray}
&&\Gamma_1[J,\Phi,\Pi]=\Gamma_1[J_{\Omega},\Phi_{\Omega},\Pi_{\Omega}]+\mbox{anomaly terms},\\
&&\Gamma_I[\Phi]=\Gamma_I[\Phi_{\Omega}]\;,\\
&&\Gamma_2[\xi_R,\xi_L,V,J,\Xi,\tilde{V},\Phi_c,\Pi_c]=\Gamma_2[1,1,V,J_{\Omega},\Xi,\tilde{V},\Phi_{\Omega c},\Pi_{\Omega c}]+\mbox{anomaly terms},\label{Gamma2-1}\\
&&S_{\mathrm{eff}}[\xi_R,\xi_L,V,J,\Xi_c,\tilde{V}_c,\Phi_c,\Pi_c]=S_{\mathrm{eff}}[1,1,V,J_{\Omega},\Xi_c,\tilde{V}_c,\Phi_{\Omega c},\Pi_{\Omega c}]+\mbox{anomaly terms}\;.\label{Seff-1}
\end{eqnarray}
Ignoring anomaly terms, Eqs.~(\ref{SeffderiveJ}) and (\ref{EMO-V-0}) become
\begin{eqnarray}
&&\hspace*{-1cm}\frac{\partial S_{\mathrm{eff}}[1,1,V,J_{\Omega},\Xi_c,\tilde{V}_c,\Phi_{\Omega c},\Pi_{\Omega c}]}{\partial J_{\Omega}^{\sigma\rho}(x)}\bigg|_{\xi_R,\xi_L,J~fix}=
N_c\bigg[\overline{\Phi_{\Omega c}^{\sigma\rho}(x,x)+\frac{\delta(0)}{4\mu^4}\slashed{\tilde{V}}_c^{\sigma\rho}(x)}\bigg]\;,
\label{SeffderiveJ-1}\\
&&\hspace*{-1cm}\frac{\partial S_{\mathrm{eff}}[1,1,V,J_{\Omega},\Xi_c,\tilde{V}_c,\Phi_{\Omega c},\Pi_{\Omega c}]}{\partial V_{\mu}^{ab}(x)}\bigg|_{\xi_R,\xi_L,J~fix}=
N_c\overline{\tilde{V}^{\mu,ba}(x)}\;.\label{EMO-V-1}
\end{eqnarray}

\section{Large $N_c$ Limit and Abelian Approximation}

In the large $N_c$ limit, all functional integrations in the last section, except those for $\xi_R$, $\xi_L$ and $V$, can be represented by their classical fields, which satisfy the stationary equation with a classical action. For simplicity, we drop the subscript $c$ in this section, then Eqs.~(\ref{Gamma2-1}) and (\ref{Seff-1}) become
\begin{eqnarray}
\Gamma_2[\xi_R,\xi_L,V,J,\Xi,\tilde{V},\Phi,\Pi]&=&\Gamma_1[J,\Phi,\Pi]\;,\label{Gamma2-Nc}\\
S_{\mathrm{eff}}[\xi_R,\xi_L,V,J,\Xi,\tilde{V},\Phi,\Pi]&=&\Gamma_2[\xi_R,\xi_L,V,J,\Xi,\tilde{V},\Phi,\Pi]\;,\nonumber\\
&=&\Gamma_2[1,1,V,J_\Omega,\Xi,\tilde{V},\Phi_{\Omega},\Pi_{\Omega}]+\mbox{anomaly terms}\;,\\
S_{\mathrm{eff}}[1,1,V,J_\Omega,\Xi,\tilde{V},\Phi_{\Omega},\Pi_{\Omega}]&=&\Gamma_1[J_{\Omega},\Phi_{\Omega},\Pi_{\Omega}]\;,
\end{eqnarray}
where we have used the property that $\Gamma_I[\Phi]$ is a quantity of order $1/N_c$. Eq.~(\ref{XictildeVc-1-0}) combined with Eqs.~(\ref{Seffdef1}) and (\ref{Gamma2def0}) now become
\begin{eqnarray}
&&e^{-i\vartheta/N_f}\mathrm{tr}_l[P_R\Phi_{\Omega}^T(x,x)]
-e^{i\vartheta/N_f}\mathrm{tr}_l[P_L\Phi_{\Omega}^T(x,x)]=0\;,
\hspace*{2cm}e^{2i\vartheta(x)}
=\frac{\mathrm{det}\bigg[\mathrm{tr}_l[P_R\Phi^T(x,x)]\bigg]}
{\mathrm{det}\bigg[\mathrm{tr}_l[P_L\Phi^T(x,x)]\bigg]}\;,
\label{constraint2}\\
&&V_{\mu}^{ab}(x)+\frac{1}{4\mu^4}\Pi_{\Omega}^{(a\xi)(b\zeta)}(x,x)
(\gamma_{\mu})^{\zeta\xi}=0\;.\label{constraint1}
\end{eqnarray}
and Eq.~(\ref{PicPhic-1-0}) becomes
\begin{eqnarray}
&&\Phi_{\Omega}^{(a\xi)(b\eta)}(x,y)+i[(i\slashed{\partial}+J_{\Omega}-\Pi_{\Omega})^{-1}]^{(b\eta)(a\xi)}(y,x)
+\frac{1}{4\mu^4}\delta(x-y)(\gamma_{\mu})^{\eta\xi}\tilde{V}^{\mu,ba}(x)=0\;,\label{EMO-Phi}\\
&&\tilde{\Xi}^{\sigma\rho}(x)\delta(x-y)+\Pi^{\sigma\rho}_{\Omega}(x,y)+{\displaystyle\sum_{n=1}^{\infty}}\int d^4x_1\cdots d^4x_nx^4x'_1\cdots d^4x'_n\frac{(-i)^{n+1}(N_cg^2)^n}{n!}
\overline{G}^{\sigma\sigma_1\cdots\sigma_n}_{\rho\rho_1\cdots\rho_n}(x,y,x_1,x'_1,\ldots,x_n,x'_n)\nonumber\\
&&\times\Phi^{\sigma_1\rho_1}_{\Omega}(x_1,x'_1)\cdots)
\Phi^{\sigma_n\rho_n}_{\Omega}(x_n,x'_n)=0\;,\label{EMO-Pi}
\end{eqnarray}
where $\tilde{V}^{\mu,ba}(x)$ and $\tilde{\Xi}^{\sigma\rho}(x)$ can be understood as effective Lagrangian multipliers, which enable the constraints (\ref{constraint1}) and (\ref{constraint2}) to hold. Note that the original discussion in Ref.~\cite{WQ0} has the constraint that\\ $\tilde{\Xi}^{\sigma\rho}(x)=\frac{\partial}{\partial\Phi_{\Omega}^{\sigma\rho}(x,x)}\int d^4y~\mathrm{tr}_{lf}\{\Xi(y)[-i\sin\vartheta(y)/N_f+\gamma_5\cos\vartheta(t)/N_f]\Phi^T_\Omega\}\bigg|_{\Xi~fix}$ only has scalar and pseudoscalar components and no vector, axial-vector or tensor components. From Eq.~(\ref{EMO-V-1}), we find $\tilde{V}^{\mu,ba}(x)$ is proportional to the l.h.s. of the EOM for a vector meson field $V_\mu^{ab}(x)$, then $\tilde{V}^{\mu,ba}(x)=0$ is the EOM of vector mesons for the effective chiral Lagrangian. We expect it could some connection with the BSE for vector mesons because the BSE is a more fundamental underlying equation describing the structure of vector mesons. For the constraint Eq.~(\ref{constraint2}), we need the coincidence limit $\Phi(x,x)$, which from Eq.~(\ref{EMO-Phi}) becomes
\begin{eqnarray}
\Phi_{\Omega}^{(a\xi)(b\eta)}(x,x)
+i[(i\slashed{\partial}+J_{\Omega}-\Pi_{\Omega})^{-1}]^{(b\eta)(a\xi)}(x,x)
+\frac{1}{4\mu^4}\delta(0)(\gamma_{\mu})^{\eta\xi}\tilde{V}^{\mu,ba}(x)=0\;,\label{EMO-Phi-1}
\end{eqnarray}
and hence we find the $\tilde{V}_\mu$ term does not make a contribution to the constraint Eq.~(\ref{constraint2}),
\begin{eqnarray}
\mathrm{tr}_l[P_{^R_L}\Phi^T(x,x)]=-i\mathrm{tr}_l\bigg[P_{^R_L}[(i\slashed{\partial}+J_{\Omega}-\Pi_{\Omega})^{-1}](x,x)\bigg]\;.\label{constraint2-1}
\end{eqnarray}
Note at the large $N_c$ limit \cite{WQ0},
\begin{eqnarray}
\overline{G}^{\sigma_1\sigma_2}_{\rho_1\rho_2}(x_1,x'_1,x_2,x'_2)=
-\frac{1}{2}G_{\mu_1\mu_2}(x_1,x_2)(\gamma^{\mu_1})_{\sigma_1\rho_2}(\gamma^{\mu_2})_{\sigma_2\rho_1}\delta(x'_1-x_2)\delta(x'_2-x_1)\;.
\end{eqnarray}
Now we take the Abelian approximation, which retains only the $n=2$ Green function $\overline{G}^{\sigma_1\sigma_2}_{\rho_1\rho_2}(x_1,x'_1,x_2,x'_2)$ and ignores all terms $n>2$ in equation Eq.~(\ref{EMO-Pi}) (If we ignore the contribution from heavy quarks, changing QCD to QED, then all the $n>2$ Green functions automatically vanish). This simplifies Eq.~(\ref{EMO-Pi}) as follows
\begin{eqnarray}
\tilde{\Xi}(x)\delta(x-y)+\Pi_{\Omega}(x,y)-\frac{1}{2}g^2N_cG_{\mu\nu}(x,y)\gamma^{\mu}\bigg[i(i\slashed{\partial}+J_{\Omega}-\Pi_{\Omega})^{-1}(x,y)+
\frac{1}{4\mu^4}\delta(x-y)\slashed{\tilde{V}}(x)\bigg]\gamma^{\nu}=0\;,
\end{eqnarray}
where we have used Eq.~(\ref{EMO-Phi}) to cancel field $\Phi_{\Omega}(x,y)$. Using the relation $G_{\mu\nu}(x,x)=G(0)g_{\mu\nu}$, we can further simplify the above equation to
\begin{eqnarray}
\bigg[\tilde{\Xi}(x)+g^2N_c\frac{G(0)}{4\mu^4}\slashed{\tilde{V}}(x)\bigg]\delta(x-y)
+\Pi_{\Omega}(x,y)-\frac{i}{2}g^2N_cG_{\mu\nu}(x,y)\gamma^{\mu}
(i\slashed{\partial}+J_{\Omega}-\Pi_{\Omega})^{-1}(x,y)\gamma^{\nu}=0\;.\label{EOM-Pi-1}
\end{eqnarray}
In this work we are only interested in the linear terms in the vector meson fields on the l.h.s. of the EOM Eq.~(\ref{EMO-V-1}), i.e., their kinetic and mass terms and not the interaction terms. For this purpose, we switch off the external source $J_{\Omega}$ and only consider the following type of solutions for $\Pi^{\sigma\rho}(x,y)$,
\begin{eqnarray}
\Pi_{\Omega}^{\sigma\rho}(x,y)\bigg|_{J_{\Omega}=0}=\delta^{\sigma\rho}\Sigma(x,y)+\int d^4z~\tilde{\Pi}_{\mu}^{\sigma\rho,ba}(x,y,z)V^{\mu,ab}(z)+O(V^2)\;,\label{PiExp}
\end{eqnarray}
where $\Sigma(x,y)$ is the quark self-energy, which satisfies the Schwinger--Dyson equation (SDE) in the rainbow approximation,
\begin{eqnarray}
\Sigma(x,y)-\frac{i}{2}g^2N_cG_{\mu\nu}(x,y)\gamma^{\mu}(i\slashed{\partial}-\Sigma)^{-1}(x,y)\gamma^{\nu}=0\;,\label{Sigma}
\end{eqnarray}
and
\begin{eqnarray}
&&\tilde{\Xi}(x)\bigg|_{J_{\Omega}=0,V_\mu=0}
=\tilde{V}_\mu(x)\bigg|_{J_{\Omega}=0,V_\mu=0}
=\theta(x)\bigg|_{J_{\Omega}=0,V_\mu=0}=0\;,\label{ConstraintVacuum}
\end{eqnarray}
where Eq.~(\ref{Sigma}) can be seen as the vacuum term of Eq.~(\ref{EOM-Pi-1}) in which both $J_{\Omega}$ and $V_\mu$ vanish, and Eq.~(\ref{ConstraintVacuum}) ensures that constraints Eqs.~(\ref{constraint2}) and (\ref{constraint1}) are naturally satisfied for the above vacuum solution Eq.~(\ref{Sigma}). The constraint (\ref{constraint1}) then is equivalent to
\begin{eqnarray}
\tilde{\Pi}_{\mu}^{(c\xi)(d\zeta),ba}(x,x,z)\bigg|_{J_{\Omega}=0,V_\mu=0}
=-\mu^4\delta(x-z)(\gamma_{\mu})^{\xi\zeta}\delta^{cb}\delta^{da}\;.\label{constraint1-1}
\end{eqnarray}
For Eq.~(\ref{EOM-Pi-1}), the terms linear in $V_\mu$ are
\begin{eqnarray}
&&\bigg[\tilde{\Xi}(x)+g^2N_c\frac{G(0)}{4\mu^4}\slashed{\tilde{V}}(x)\bigg]\delta(x-y)+
\int d^4z~\tilde{\Pi}_{\lambda}^{ba}(x,y,z)V^{\lambda,ab}(z)\label{EOM-Pi-2}\\
&&-\frac{i}{2}g^2N_cG_{\mu\nu}(x,y)\int d^4x'd^4y'd^4z\gamma^{\mu}
(i\slashed{\partial}+J_{\Omega}-\Sigma)^{-1}(x,x')\tilde{\Pi}_{\lambda}^{ba}(x',y',z)V^{\lambda,ab}(z)(i\slashed{\partial}+J_{\Omega}-\Sigma)^{-1}(y',y)\gamma^{\nu}=0\;.\nonumber
\end{eqnarray}
In Eq.~(\ref{EOM-Pi-2}), the two terms involving the effective Lagrangian multipliers $\tilde{\Xi}(x)$ and $\slashed{\tilde{V}}(x)$ can be treated as inhomogeneous terms of the equation. Ignoring these inhomogeneous terms and switching off the external source $J_{\Omega}$, we obtain
\begin{eqnarray}
&&\int d^4z~\tilde{\Pi}_{\mathrm{BS},\lambda}^{ba}(x,y,z)V^{\lambda,ab}(z)\;,~~~\label{BS}\\
&&-\frac{i}{2}g^2N_cG_{\mu\nu}(x,y)\int d^4x'd^4y'\gamma^{\mu}
(i\slashed{\partial}-\Sigma)^{-1}(x,x')\int d^4z~\tilde{\Pi}_{\mathrm{BS},\lambda}^{ba}(x',y',z)V^{\lambda,ab}(z)
(i\slashed{\partial}-\Sigma)^{-1}(y',y)\gamma^{\nu}=0\;,\nonumber
\end{eqnarray}
which is the standard homogeneous BSE in the ladder approximation for the bound state with amplitude $\tilde{\Pi}_{\mathrm{BS},\lambda}^{ba}(x,y,z)$. Here we use a subscript BS to denote that it is a BS amplitude. For its vector meson solutions, Ref.~\cite{BSvector} presents a detailed analysis, for which Eq.~(\ref{BS}) is just the starting equation (4) of Ref.~\cite{BSvector}.

Now, if our $\tilde{\Pi}_{\lambda}^{ba}(x,y,z)$ in Eq.~(\ref{EOM-Pi-2}) happens to be the BS amplitude $\tilde{\Pi}_{\mathrm{BS},\lambda}^{ba}(x,y,z)$ given by Eq.~(\ref{BS}) that fix the vector meson structure, we expect these properties will have some impact on the corresponding vector field $V_\mu$. The action on the vector field can be obtained by applying Eq.~(\ref{BS}) to Eq.~(\ref{EOM-Pi-2}),
\begin{eqnarray}
&&\tilde{\Xi}(x)\bigg|_{J_{\Omega}=0,\mathrm{linear~in~}V}=0\;,\label{tildeXivector}\\
&&\slashed{\tilde{V}}(x)\bigg|_{J_{\Omega}=0,\mathrm{linear~in~}V}=0\;,\label{EOMvector}
\end{eqnarray}
where the interpretation of Eq.~(\ref{tildeXivector}) is that the effective Lagrangian multiplier $\tilde{\Xi}$ has no linearly dependent $V$ term. For Eq.~(\ref{EOMvector}) combined with Eq.~(\ref{EMO-V-1}) (the average operation in the formula has no effect at large $N_c$ limit), the implication is that the vector meson field satisfies its own EOM. Alternatively, in momentum space, the vector meson is on its mass shell. This elucidates the fundamental nature of the vector meson determined by Eq.~(\ref{BS}) in regard to the phenomenological EOM of the vector field.  We can state this result in a different way. We demand that the vector meson field satisfy its own EOM and hence from Eq.~(\ref{EMO-V-1}) implies Eq.~(\ref{EOMvector}). If we substitute this equation into Eq.~(\ref{EOM-Pi-2}), then its vector part becomes BSE (\ref{BS}); i.e., the EOM of vector meson field is equivalent to its BSE. Therefore, we have established a connection between the BSE for the bound state at the underlying level of QCD and the EOM at the phenomenological effective chiral Lagrangian level for the vector mesons. In our formalism, the BSE can lead to the EOM, and vice versa. Indeed, if we require $\mu$ independence in Eq.~(\ref{EOM-Pi-2}), we get Eq.~(\ref{EOMvector}); i.e., the resulting $\mu$ independence is a key requirement to relate the fundamental BSE (\ref{BS}) with the phenomenological EOM (\ref{EOMvector}).

At the present stage, because we only considered the terms linear in vector field in the expansion of Eq.~(\ref{PiExp}), this implies that the resulting EOM only includes kinetic and mass terms and no interaction term. This arrangement matches the general large $N_c$ approximation that we have taken and the corresponding general result given in Ref.~\cite{largeN} that the mesons in this limit are non-interacting particles. It is a miracle here that these two important equations obtained at different scales (quark gluon scale and meson scale) are related so closely in our formalism. Note here the mass of the vector meson is determined through BSE (\ref{BS}), which has already discussed in detail in Ref.~\cite{BSvector}. Despite the arbitrary parameter $\mu$ in our formalism, the mass itself and the amplitude of the vector mesons are independent of this parameter.

Although from BSE we can obtain EOM, the question is whether the mass fixed in BSE is the same as that appearing in EOM? The answer is yes. To confirm this, we need to compute the 2-point vertex of the vector meson fields or the inverse of the vector meson propagator, which in fact is the operator in front of EOM of free meson field,
 \begin{eqnarray}
 &&\hspace*{-1cm}\frac{\partial^2 S_{\mathrm{eff}}[1,1,V,J_\Omega,\Xi,\tilde{V},\Phi_{\Omega},\Pi_{\Omega}]}{\partial V^{\mu,ab}(x)\partial V^{\nu,cd}(x')}\bigg|_{J_{\Omega}=0}
  =\int d^4yd^4zd^4y'd^4z'\bigg[\frac{\partial^2 S_{\mathrm{eff}}}{\partial\Pi_{\Omega}^{\sigma\rho}(y,z)\partial\Pi_{\Omega}^{\sigma'\rho'}(y',z')}
 \frac{\partial\Pi_{\Omega}^{\sigma\rho}(y,z)}{\partial V^{\mu,ab}(x)}\frac{\partial\Pi_{\Omega}^{\sigma'\rho'}(y',z')}{\partial V^{\nu,cd}(x')}
  \nonumber\\
 &&\hspace*{0cm}+\frac{\partial^2 S_{\mathrm{eff}}}{\partial\Phi_{\Omega}^{\sigma\rho}(y,z)\partial\Phi_{\Omega}^{\sigma'\rho'}(y',z')}
 \frac{\partial\Phi_{\Omega}^{\sigma\rho}(y,z)}{\partial V^{\mu,ab}(x)}\frac{\partial\Phi_{\Omega}^{\sigma'\rho'}(y',z')}{\partial V^{\nu,cd}(x')}
+2\frac{\partial^2 S_{\mathrm{eff}}}
{\partial\Phi_{\Omega}^{\sigma\rho}(y,z)\partial\Pi_{\Omega}^{\sigma'\rho'}(y',z')}
 \frac{\partial\Phi_{\Omega}^{\sigma\rho}(y,z)}{\partial V^{\mu,ab}(x)}\frac{\partial\Pi_{\Omega}^{\sigma'\rho'}(y',z')}{\partial V^{\nu,cd}(x')}\bigg]\nonumber\\
  &&\hspace*{-1cm}=-iN_c\int d^4y_1d^4z_1\bigg\{ \tilde{\Pi}_{\nu}^{dc}(y_1,z_1,x')
-\frac{i}{2}N_cg^2G_{\mu_1\mu_2}(y_1,z_1)
  \bigg[\gamma^{\mu_1}(i\slashed{\partial}\!-\!\Sigma)^{-1}\tilde{\Pi}_{\nu}^{dc}(x')(i\slashed{\partial}\!-\!\Sigma)^{-1}\gamma^{\mu_2}\bigg](y_1,z_1)\bigg\}^{\sigma_1\sigma_2}\nonumber\\
  &&\times\bigg[(i\slashed{\partial}\!-\!\Sigma)^{-1}\tilde{\Pi}_{\mu}^{ba}(x)(i\slashed{\partial}\!-\!\Sigma)^{-1}\bigg]^{\sigma_2\sigma_1}(z_1,y_1)\;.\label{2pointvertex}
  \end{eqnarray}
If we identify the $\tilde{\Pi}_{\nu}^{dc}$ appearing in Eq.~(\ref{2pointvertex}) with the BS amplitude $\tilde{\Pi}_{\mathrm{BS},\nu}^{dc}$ determined by BSE (\ref{BS}), then the above 2-point vertex vanishes. This just shows that the mass term in the 2-point vertex is the same as that fixed by BSE, as BSE fixes the momentum space 2-point vertex onto the mass shell.
\section{Include scalar and axial-vector mesons}

With the exception of vector meson fields, we now include the scalar and axial-vector meson fields. We start from the QCD generating functional Eq.~(\ref{ZxiRL}), and functionally insert an alternative constant integration,
\begin{eqnarray}
&&\int {\cal D}\phi~\delta\bigg[\phi^{(a\xi)(b\zeta)}(x)+\frac{1}{\mu^4}
\{[e^{-\frac{i\vartheta(x)}{2N_f}}\xi_L(x)P_R+e^{\frac{i\vartheta(x)}{2N_f}}\xi_R(x)P_L]\Pi(x,x)
[e^{-\frac{i\vartheta(x)}{2N_f}}\xi_R^\dag(x)P_R+e^{\frac{i\vartheta(x)}{2N_f}}
\xi_L^\dag(x)P_L]\}^{(a\xi')(b\zeta')}P^{\xi'\zeta',\xi\zeta}\bigg]\nonumber\\
\label{phiinsert}
\end{eqnarray}
where $\vartheta$ is determined by Eq.~(\ref{constraint2}), $P^{\xi'\zeta',\xi\zeta}$ is the projection operator, which projects a general four by four matrix into its scalar, vector, and axial-vector subspaces,
\begin{eqnarray}
P^{\xi'\zeta',\xi\zeta}=\frac{1}{4}\bigg[\delta^{\zeta'\xi'}\delta^{\xi\zeta}
+(\gamma_{\mu})^{\zeta'\xi'}(\gamma^{\mu})^{\xi\zeta}
-(\gamma_{\mu}\gamma_5)^{\zeta'\xi'}(\gamma^{\mu}\gamma_5)^{\xi\zeta}
\bigg]\;.\label{Projector}
\end{eqnarray}
Note the completion relation
\begin{eqnarray}
\delta^{\xi\xi'}\delta^{\zeta'\zeta}=\frac{1}{4}(\gamma_5)^{\zeta'\xi'}(\gamma_5)^{\xi\zeta}
+\frac{1}{8}(\sigma_{\mu\nu})^{\zeta'\xi'}(\sigma^{\mu\nu})^{\xi\zeta}+P^{\xi'\zeta',\xi\zeta}\;.
\end{eqnarray}
Exponentiating the delta function by introducing functional integration over another field $\tilde{\phi}^{\rho\sigma}(x)$ and integrating out the fields $\Pi$ and $\Phi$, $Z[J]$ can be rearranged to give
\begin{eqnarray}
Z[J]=\int{\cal D}\xi_R{\cal D}\xi_L{\cal D}\Xi{\cal D}\phi{\cal D}
\tilde{\phi}~\delta(\xi_R^\dag\xi_R-1)\delta(\xi_L^\dag\xi_L-1)
\delta(\mathrm{det}\xi_R-\mathrm{det}\xi_L)
e^{i\Gamma_2[\xi_R,\xi_L,\phi,J,\Xi,\tilde{\phi},\Phi_c,\Pi_c]}\;,
\end{eqnarray}
 where
 \begin{eqnarray}
&& e^{i\Gamma_2[\xi_R,\xi_L,\phi,J,\Xi,\tilde{\phi},\Phi_c,\Pi_c]}=\int{\cal D}\Phi{\cal D}\Pi~\exp\bigg\{i\Gamma_1[J,\Phi,\Pi]+i\Gamma_I[\Phi]\nonumber\\
&&\hspace*{0cm}+iN_c\int d^4x\bigg[\phi^{(a\xi)(b\zeta)}(x)+\frac{1}{\mu^4}
\{[e^{-\frac{i\vartheta(x)}{2N_f}}\xi_L(x)P_R+e^{\frac{i\vartheta(x)}{2N_f}}\xi_R(x)P_L]\Pi(x,x)
[e^{-\frac{i\vartheta(x)}{2N_f}}\xi_R^\dag(x)P_R+e^{\frac{i\vartheta(x)}{2N_f}}\xi_L^\dag(x)P_L]\}^{(a\xi')(b\zeta')}\nonumber\\
&&\times P^{\xi'\zeta',\xi\zeta}\bigg]\tilde{\phi}^{(b\zeta)(a\xi)}(x)+iN_c\int d^4x\mathrm{tr}_f[\Xi\{e^{-i\vartheta/N_f}\xi_R\mathrm{tr}_l(P_R\Phi^T)\xi_L^\dag-e^{i\vartheta/N_f}\xi_L\mathrm{tr}_l(P_L\Phi^T)\xi_R^\dag\}]\bigg\}\;.\label{Gamma2def}
\end{eqnarray}
The $\Pi_c$ and $\Phi_c$ fields still satisfy Eq.~(\ref{PicPhic0}), but $\cdots$ in Eq.~(\ref{PicPhic0}) now is the exponent to the exponential in Eq.~(\ref{Gamma2def}). $\Pi_c$ and $\Phi_c$ satisfy the following equations
\begin{eqnarray}
\frac{\partial\Gamma_2[\xi_R,\xi_L,\phi,J,\Xi,\tilde{\phi},\Phi_c,\Pi_c]}{\partial\Pi_c}=0,\hspace*{2cm}\frac{\partial\Gamma_2[\xi_R,\xi_L,\phi,J,\Xi,\tilde{\phi},\Phi_c,\Pi_c]}{\partial\Phi_c}=0,
\label{PicPhic-1}
\end{eqnarray}
We formally finish the integration over fields $\tilde{\phi}$ and $\Xi$,
\begin{eqnarray}
Z[J]=\int{\cal D}\xi_R{\cal D}\xi_L{\cal D}\phi
\delta(\xi_R^\dag\xi_R-1)\delta(\xi_L^\dag\xi_L-1)\delta(\mathrm{det}\xi_R-\mathrm{det}\xi_L)
e^{iS_{\mathrm{eff}}[\xi_R,\xi_L,\phi,J,\Xi_c,\tilde{\phi}_c,\Phi_c,\Pi_c]}\;,
\end{eqnarray}
where
\begin{eqnarray}
e^{iS_{\mathrm{eff}}[\xi_R,\xi_L,\phi,J,\Xi_c,\tilde{\phi}_c,\Phi_c,\Pi_c]}=\int{\cal D}\Xi
{\cal D}\tilde{\phi}~e^{i\Gamma_2[\xi_R,\xi_L,\phi,J,\Xi,\tilde{\phi},\Phi_c,\Pi_c]}\;,
\end{eqnarray}
$\Xi_c(x)$ and $\tilde{\phi}(x)$ are
\begin{eqnarray}
\Xi_c=\frac{\int{\cal D}\Xi{\cal D}\tilde{\phi}~\Xi
~e^{i\Gamma_2[\xi_R,\xi_L,\phi,J,\Xi,\tilde{\phi},\Phi_c,\Pi_c]}}{\int{\cal D}\Xi{\cal D}\tilde{\phi}
~e^{i\Gamma_2[\xi_R,\xi_L,\phi,J,\Xi,\tilde{\phi},\Phi_c,\Pi_c]}},\hspace*{1cm}
\tilde{\phi}_c=\frac{\int{\cal D}\Xi{\cal D}\tilde{\phi}~\tilde{\phi}
~e^{i\Gamma_2[\xi_R,\xi_L,\phi,J,\Xi,\tilde{\phi},\Phi_c,\Pi_c]}}{\int{\cal D}\Xi{\cal D}\tilde{\phi}
~e^{i\Gamma_2[\xi_R,\xi_L,\phi,J,\Xi,\tilde{\phi},\Phi_c,\Pi_c]}},\label{Xicphic}
\end{eqnarray}
which satisfy
\begin{eqnarray}
\frac{\partial S_{\mathrm{eff}}[\xi_R,\xi_L,\phi,J,\Xi_c,\tilde{\phi}_c,\Phi_c,\Pi_c]}
{\partial\Xi_c}=0,\hspace*{2cm}
\frac{\partial S_{\mathrm{eff}}[\xi_R,\xi_L,\phi,J,\Xi_c,\tilde{\phi}_c,\Phi_c,\Pi_c]}
{\partial\tilde{\phi}_c}=0.\label{Xicphic-1}
\end{eqnarray}
Aided by Eqs.~(\ref{Gamma2def}), (\ref{PicPhic-1}) and (\ref{Xicphic-1}), one can show that
\begin{eqnarray}
&&\frac{dS_{\mathrm{eff}}[\xi_R,\xi_L,\phi,J,\Xi_c,\tilde{\phi}_c,\Phi_c,\Pi_c]}{d J^{\sigma\rho}(x)}\bigg|_{\xi_R,\xi_L,\phi~\mathrm{fix}}\nonumber\\
&&=N_c\bigg[\overline{\Phi_c^{\sigma\rho}(x,x)+\frac{\delta(0)}{\mu^4}
\{[e^{\frac{i\vartheta(x)}{2N_f}}\xi_R^\dag(x)P_R
+e^{-\frac{i\vartheta(x)}{2N_f}}\xi_L^\dag(x)P_L]
P\tilde{\phi}(x)[e^{\frac{i\vartheta(x)}{2N_f}}\xi_L(x)P_R
+e^{-\frac{i\vartheta(x)}{2N_f}}\xi_R(x)P_L]\}^{\sigma\rho}}\bigg]\;,\label{SeffJ}
\end{eqnarray}
which is useful in determining the LECs of the effective chiral Lagrangian. We use a bar to denote the functional average over fields $\Xi$ and $\tilde{\phi}$,
\begin{eqnarray}
\overline{O(x)}\equiv N_c\frac{\int{\cal D}\Xi{\cal D}\tilde{\phi}~O(x)
~\exp\{i\Gamma_2[\xi_R,\xi_L,\phi,J,\Xi,\tilde{\phi},\Phi_c,\Pi_c]\}}
{\int{\cal D}\Xi{\cal D}\tilde{\phi}
~\exp\{i\Gamma_2[\xi_R,\xi_L,\phi,J,\Xi,\tilde{\phi},\Phi_c,\Pi_c]\}}\;.
\end{eqnarray}
One can similarly find
\begin{eqnarray}
\frac{dS_{\mathrm{eff}}[\xi_R,\xi_L,\phi,J,\Xi_c,\tilde{\phi}_c,\Phi_c,\Pi_c]}
{d\phi^{(a\xi)(b\zeta)}(x)}\bigg|_{\xi_R,\xi_L,\phi~\mathrm{fix}}
=N_cP^{\xi\zeta,\xi'\zeta'}\overline{\tilde{\phi}_c^{(b\zeta')(a\xi')}(x)}\;.
\label{EOMphi}
\end{eqnarray}
Defining the rotated source and fields as follows,
\begin{eqnarray}
&&J_{\Omega}(x)=[e^{i\vartheta(x)/N_f}\xi_L(x)P_R+e^{-i\vartheta(x)/N_f}\xi_R(x)P_L][J(x)+i\slashed{\partial}][e^{i\vartheta(x)/N_f}\xi_R^\dag(x)P_R+e^{-i\vartheta(x)/N_f}\xi_L^\dag(x)P_L],\nonumber\\
&&\Phi^T_{\Omega}(x,y)=[e^{-i\vartheta(x)/N_f}\xi_R(x)P_R+e^{i\vartheta(x)/N_f}\xi_L(x)P_L]\Phi^T(x,y)[e^{-i\vartheta(y)/N_f}\xi_L^\dag(y)P_R+e^{i\vartheta(y)/N_f}\xi_R^\dag(y)P_L],\nonumber\\
&&\Pi_{\Omega}(x,y)=[e^{i\vartheta(x)/N_f}\xi_L(x)P_R+e^{-i\vartheta(x)/N_f}\xi_R(x)P_L]\Pi(x,y)[e^{i\vartheta(y)/N_f}\xi_R^\dag(y)P_R+e^{-i\vartheta(y)/N_f}\xi_L^\dag(y)P_L],\label{rotation1}
\end{eqnarray}
the fields $\theta$,$\Xi$,$\phi$,$\tilde{\phi}$ remain unchanged. There is no explicit $\xi_L$, $\xi_R$-dependence in Eq.~(\ref{Gamma2def}) after rotation as all the $\xi_R$ and $\xi_L$ dependence is absorbed into the variables distinguished by subscript $\Omega$.
\begin{eqnarray}
&&\Gamma_2[\xi_R,\xi_L,\phi,J,\Xi,\tilde{\phi},\Phi_c,\Pi_c]=\Gamma_2[1,1,\phi,J_{\Omega},\Xi,\tilde{\phi},\Phi_{\Omega c},\Pi_{\Omega c}]+\mbox{anomaly terms},\label{Gamma2rot}\\
&&S_{\mathrm{eff}}[\xi_R,\xi_L,\phi,J,\Xi_c,\tilde{\phi}_c,\Phi_c,\Pi_c]=S_{\mathrm{eff}}[1,1,\phi,J_{\Omega},\Xi_c,\tilde{\phi}_c,\Phi_{\Omega c},\Pi_{\Omega c}]
+\mbox{anomaly terms},\label{Seffrot}
\end{eqnarray}
where
\begin{eqnarray}
&&e^{i\Gamma_2[1,1,\phi,J_\Omega,\Xi,\tilde{\phi},\Phi_{\Omega c},\Pi_{\Omega c}]}=\int{\cal D}\Phi_{\Omega}{\cal D}\Pi_{\Omega}\exp\bigg\{i\Gamma_1[J_{\Omega},\Phi_{\Omega},\Pi_{\Omega}]+\Gamma_I[\Phi_\Omega]\\
&&+iN_c\int d^4x\bigg[\phi^{(a\xi)(b\zeta)}+\frac{1}{\mu^4}
\Pi_{\Omega}^{(a\xi')(b\zeta')}P^{\xi'\zeta',\xi\zeta}\bigg]\tilde{\phi}^{(b\zeta)(a\xi)}
+N_c\int d^4x\mathrm{tr}_f\bigg[\Xi[\mathrm{tr}_l(\gamma_5\Phi_{\Omega}^T)]\bigg]\bigg\}
\;.\nonumber
\end{eqnarray}
Because the Jacobi terms coming from $\Phi\rightarrow\Phi_{\Omega}$ and $\Pi\rightarrow\Pi_{\Omega}$ cancel, the functional integration measure does not change, i.e., ${\cal D}\Phi{\cal D}\Pi={\cal D}\Phi_{\Omega}{\cal D}\Pi_{\Omega}$. Equations Eqs.~(\ref{Gamma2def}), (\ref{PicPhic-1}),(\ref{Xicphic}),(\ref{Xicphic-1}) and (\ref{SeffJ}) are the same as before except one must change all quantities with subscript $\Omega$ and add an anomaly term into the numerator of Eqs.~(\ref{PicPhic-1}), (\ref{Xicphic-1}) and (\ref{SeffJ}). In particular, by ignoring the anomaly, Eq.~(\ref{SeffJ}) on the rotated basis is
\begin{eqnarray}
\frac{dS_{\mathrm{eff}}[1,1,\phi,J_{\Omega},\Xi_c,\tilde{\phi}_c,\Phi_{\Omega c},\Pi_{\Omega c}]}{d J^{(a\xi)(b\zeta)}_{\Omega}(x)}\bigg|_{\xi_R,\xi_L,\phi~\mathrm{fix, no anomaly}}=
N_c\bigg[\overline{\Phi_{\Omega c}^{(a\xi)(b\zeta)}(x,x)+\frac{\delta(0)}{\mu^4}P^{\xi\zeta,\xi'\zeta'}\tilde{\phi}^{(a\xi')(b\zeta')}_c(x)}\bigg]\;.\label{SeffJ-1}
\end{eqnarray}
Equation (\ref{EOMphi}) becomes
\begin{eqnarray}
\frac{dS_{\mathrm{eff}}[1,1,\phi,J_{\Omega},\Xi_c,\tilde{\phi}_c,\Phi_{\Omega c},\Pi_{\Omega c}]}{d\phi^{(a\xi)(b\zeta)}(x)}\bigg|_{\xi_R,\xi_L,\phi~\mathrm{fix}}=N_cP^{\xi\zeta,\xi'\zeta'}\overline{\tilde{\phi}_c^{(b\zeta')(a\xi')}(x)}\;.
\label{EOMphi-1}
\end{eqnarray}
At the large $N_c$ limit,
\begin{eqnarray}
S_{\mathrm{eff}}[\xi_R,\xi_L,\phi,J,\Xi,\tilde{\phi},\Phi,\Pi]&=&\Gamma_2[\xi_R,\xi_L,\phi,J,\Xi,\tilde{\phi},\Phi,\Pi]\;,\nonumber\\
&=&\Gamma_2[1,1,\phi,J_\Omega,\Xi,\tilde{\phi},\Phi_{\Omega},\Pi_{\Omega}]+\mbox{anomaly terms}\;,\\
\Gamma_2[1,1,\phi,J_\Omega,\Xi,\tilde{\phi},\Phi_{\Omega},\Pi_{\Omega}]&=&\Gamma_1[J_{\Omega},\Phi_{\Omega},\Pi_{\Omega}]\;,\\
S_{\mathrm{eff}}[1,1,\phi,J_\Omega,\Xi,\tilde{\phi},\Phi_{\Omega},\Pi_{\Omega}]&=&\Gamma_1[J_{\Omega},\Phi_{\Omega},\Pi_{\Omega}]\;,
\end{eqnarray}
and
\begin{eqnarray}
&&\mathrm{tr}_l(\gamma_5\Phi_{\Omega}^T)=0\;,\label{constraint2new}\\
&&\phi^{(a\xi)(b\zeta)}+\frac{1}{\mu^4}\Pi_{\Omega}^{(a\xi')(b\zeta')}P^{\xi'\zeta',\xi\zeta}=0\;,\label{constraint1new}\\
&&\Phi_{\Omega}^{(a\xi)(b\zeta)}(x,y)+i[(i\slashed{\partial}+J_{\Omega}-\Pi_{\Omega})^{-1}]^{(b\zeta)(a\xi)}(y,x)
+\frac{1}{\mu^4}\delta(x-y)P^{\zeta\xi,\zeta'\xi'}\tilde{\phi}^{(b\zeta')(a\xi')}(x)=0\;,\label{EMO-Phinew}\\
&&\frac{\partial S_{\mathrm{eff}}[1,1,\phi,J_\Omega,\Xi,\tilde{\phi},\Phi_{\Omega},\Pi_{\Omega}]}{\partial\phi^{\sigma\rho}(x)}\bigg|_{\xi_R,\xi_L,J~fix}=
N_c\tilde{\phi}^{\rho\sigma}(x)\;.\label{EMO-Vnew}
\end{eqnarray}
Here, we only list the equations modified by introducing the additional meson fields; Eqs.~(\ref{Gamma1def}) and (\ref{EMO-Pi}) remain the same as the pure vector meson field case.
Equation (\ref{constraint2new}) is a modified version of Eq.~(\ref{constraint2}); the change occurs because, in Eqs.~(\ref{phiinsert}) and (\ref{rotation1}), we have included an extra $U(1)$ rotation $e^{\frac{i\vartheta(x)}{2N_f}}$. Equation (\ref{constraint2new}) implies that in the present formalism $\Phi_{\Omega}^T(x,x)$ does not have a pseudoscalar component (therefore we have no need to consider a pseudoscalar component in the $\phi$ field), as this component is already extracted as a pseudoscalar meson field $U(x)=\xi_L^\dag(x)\xi_R(x)$. Taking the Abelian approximation, Eq.~(\ref{EMO-Pi}) becomes
\begin{eqnarray}
\tilde{\Xi}(x)\delta(x-y)+\Pi_{\Omega}(x,y)-\frac{1}{2}g^2N_cG_{\mu\nu}(x,y)
\gamma^{\mu}\bigg[i(i\slashed{\partial}+J_{\Omega}-\Pi_{\Omega})^{-1}(x,y)
+\frac{1}{\mu^4}\delta(x-y)P\tilde{\phi}(x)\bigg]\gamma^{\nu}=0\;,
\end{eqnarray}
where we have used Eq.~(\ref{EMO-Phinew}) to cancel field $\Phi_{\Omega}(x,y)$, and $\tilde{\Xi}(x)$ is the effective Lagrangian multiplier for the pseudoscalar meson field $\tilde{\Xi}^{(a\xi)(b\zeta)}(x)=\Xi^{ab}(x)(\gamma_5)^{\xi\zeta}$, which only has a $\gamma_5$ component. We can further simplify the above equation to
\begin{eqnarray}
\bigg[\tilde{\Xi}(x)-\frac{1}{2}g^2N_c\frac{G(0)}{\mu^4}\gamma^{\mu}P\tilde{\phi}(x)\gamma_{\mu}\bigg]\delta(x-y)+\Pi_{\Omega}(x,y)-\frac{i}{2}g^2N_cG_{\mu\nu}(x,y)\gamma^{\mu}
(i\slashed{\partial}+J_{\Omega}-\Pi_{\Omega})^{-1}(x,y)\gamma^{\nu}=0\;.\label{EOM-Pi-1new}
\end{eqnarray}
Switching off external source $J_{\Omega}$ and considering only the following type of solution $\Pi^{\sigma\rho}(x,y)$, then
\begin{eqnarray}
\Pi_{\Omega}^{\sigma\rho}(x,y)\bigg|_{J_{\Omega}=0}=\delta^{\sigma\rho}\Sigma(x,y)
+\int d^4z~\tilde{\Pi}^{\sigma\rho,\sigma'\rho'}(x,y,z)[P\phi(z)]^{\sigma'\rho'}
+O(\phi^2)\;.\label{PiExpnew}
\end{eqnarray}
The constraint (\ref{constraint1new}) then is equivalent to
\begin{eqnarray}
\tilde{\Pi}^{\sigma\rho,\sigma'\rho'}(x,x,z)\bigg|_{J_{\Omega}=0,\phi=0}
=-\mu^4\delta(x-z)\delta^{\sigma\sigma'}\delta^{\rho\rho'}\;.\label{constraint1-1new}
\end{eqnarray}

For (\ref{EOM-Pi-1new}), the terms linear in $\phi$ are
\begin{eqnarray}
&&\bigg[\tilde{\Xi}(x)-\frac{1}{2}g^2N_c\frac{G(0)}{\mu^4}
\gamma_{\mu}P\tilde{\phi}(x)\gamma^{\mu}\bigg]\delta(x-y)
+\int d^4z~\tilde{\Pi}_{\lambda}^{\sigma'\rho'}(x,y,z)
[P\phi(z)]^{\sigma'\rho'}\label{EOM-Pi-2new}\\
&&-\frac{i}{2}g^2N_cG_{\mu\nu}(x,y)\int d^4x'd^4y'd^4z\gamma^{\mu}
(i\slashed{\partial}+J_{\Omega}-\Sigma)^{-1}(x,x')\tilde{\Pi}^{\sigma'\rho'}(x',y',z)
[P\phi(z)]^{\sigma'\rho'}
(i\slashed{\partial}+J_{\Omega}-\Sigma)^{-1}(y',y)\gamma^{\nu}=0\;.\nonumber
\end{eqnarray}
In Eq.~(\ref{EOM-Pi-2new}), the two terms involving effective Lagrangian multipliers $\tilde{\Xi}(x)$ and $\gamma_{\mu}P\tilde{\phi}(x)\gamma^{\mu}$ can be treated as inhomogeneous terms of the equation. Ignoring these inhomogeneous terms and switching off the external source $J_{\Omega}$, we obtain the following equation,
\begin{eqnarray}
&&\int d^4z~\tilde{\Pi}_{\mathrm{BS}}^{\sigma'\rho'}(x,y,z)[P\phi(z)]^{\sigma'\rho'}~~~\label{BSnew}\\
&&-\frac{i}{2}g^2N_cG_{\mu\nu}(x,y)\int d^4x'd^4y'\gamma^{\mu}(i\slashed{\partial}
-\Sigma)^{-1}(x,x')\int d^4z~\tilde{\Pi}_{\mathrm{BS}}^{\sigma'\rho'}(x',y',z)
[P\phi(z)]^{\sigma'\rho'}(i\slashed{\partial}-\Sigma)^{-1}(y',y)\gamma^{\nu}=0\;,\nonumber
\end{eqnarray}
which is also the standard homogeneous BSE in the ladder approximation for the bound state with amplitude $\tilde{\Pi}_{\mathrm{BS}}^{\sigma'\rho'}(x,y,z)$.

Whereas Ref.~\cite{BSvector} describes vector mesons, Ref.~\cite{BSsolution1} describes scalar and axial-vector mesons.
If our $\tilde{\Pi}^{\sigma'\rho'}(x,y,z)$ in Eq.~(\ref{EOM-Pi-2new}) is just the BS amplitude $\tilde{\Pi}_{\mathrm{BS}}^{\sigma'\rho'}(x,y,z)$ given by Eq.~(\ref{BSnew}), then Eq.~(\ref{EOM-Pi-2new}) for this BS amplitude further implies that
\begin{eqnarray}
&&\tilde{\Xi}(x)=0\;,\label{tildeXiaxilvector}\\
&&\gamma_{\mu}P\tilde{\phi}(x)\gamma^{\mu}=0\;.\label{EOMphi-1}
\end{eqnarray}
Note that for the Lorentz index structures, $\tilde{\Xi}$ only has a pseudoscalar part, whereas $\gamma_{\mu}P\tilde{\phi}(x)\gamma^{\mu}$ only has a scalar, vector, and axial-vector parts, with each independent part vanishing separately. The interpretation of Eq.~(\ref{tildeXiaxilvector}) is that the effective Lagrangian multiplier $\tilde{\Xi}$ has no linearly dependent $\phi$ (which includes scalar, vector, and axial-vector) term. For Eq.~(\ref{EOMphi-1}), in combination with Eq.~(\ref{EMO-Vnew}), Eq.~(\ref{EOMphi-1}) implies that the scalar, vector, and axial-vector meson fields satisfying their own EOM. As done in Eq.~(\ref{2pointvertex}) in the last section, we can prove by computing 2-point vertices that the masses appearing in the EOM are the same as those determined by the BSE (\ref{BSnew}).


\section{Summary and Discussion}

We reinvestigated the derivation of the effective vector meson from QCD to recognize that it can reproduce both BSE and EOM for the vector meson, and we show that the mass appearing in the EOM is the same as that determined by the BSE. The computation is done in the large $N_c$ limit and with Abelian approximation. We know that in the large $N_c$ limit mesons are free particles and this makes the discussion much simpler. The Abelian approximation makes our resulting SDE (\ref{Sigma}) equivalent to that in the standard rainbow approximation, and the BSE (\ref{BS}) equivalent to that in the conventional ladder approximation. We can easily go beyond the Abelian approximation to reformulate our formalism, but that requires a SDE and a BSE that extend beyond the traditional rainbow and ladder approximations. In Ref.~\cite{SDE}, we have made preliminary investigation into the effects of SDE beyond the rainbow approximation.

We have generalized the formalism for vector mesons to further include scalar and axial-vector mesons, with qualitative results similar to those for vector mesons. One may ask why we only consider scalar and axial-vector mesons. In general, the projection operator (\ref{Projector}) can further include pseudoscalar and antisymmetric tensor parts, which corresponds to introducing pseudoscalar and antisymmetric tensor mesons. As noted for pseudoscalar mesons, we have already introduced this degree of freedom as $\xi_R$ and $\xi_L$ through the chiral rotation and there is no need to reintroduce them. Even if one does introduce them into the theory, by considering the pseudoscalar part $\frac{1}{4}(\gamma_5)^{\zeta'\xi'}(\gamma_5)^{\xi\zeta}$ for the extra pseudoscalar meson field in the projection operator (\ref{Projector}), this pseudoscalar part is trivial and will play no role in our theory because of constraint (\ref{constraint2new}). This result can be seen as follows; taking the coincidence limit of Eq.~(\ref{EMO-Phinew}), then constraints (\ref{constraint2new}) and (\ref{EOMphi-1}) imply the pseudoscalar component of $(i\slashed{\partial}+J_{\Omega}-\Pi_{\Omega})^{-1}(x,x)$ is zero. Substituting this result into the coincidence limit of Eq.~(\ref{EOM-Pi-1new}) and combining it with Eqs.~(\ref{tildeXiaxilvector}) and (\ref{EOMphi-1}), the pseudoscalar part of $\Pi_{\Omega}(x,x)$ vanishes, which from constraint (\ref{constraint1new}) implies the pseudoscalar part of introduced field $\phi(x)$ is zero. For tensor mesons, Refs.~\cite{tensor1,tensor2} have proved that at the effective chiral Lagrangian level, the antisymmetric tensor fields description of mesons is equivalent to a vector meson fields description. With this equivalence, one can change our resultant vector meson chiral Lagrangian given by Ref.~\cite{WQ00vector} into its antisymmetric tensor version. As we have already introduced a vector meson field into our theory, there is no need to consider a corresponding antisymmetric tensor part. If the antisymmetric tensor fields is introduced, then by consider in antisymmetric tensor part $\frac{1}{8}(\sigma_{\mu\nu})^{\zeta'\xi'} (\sigma^{\mu\nu})^{\xi\zeta}$ in the projection operator (\ref{Projector}), then one finds that for the antisymmetric tensor part, the EOM of Eq.~(\ref{EOMphi-1}) is an identity due to fact
of $\gamma_\mu\sigma^{\mu'\nu'}\gamma^\mu=0$, i.e., it is just trivially satisfied without any relation to its underlying BSE. This is an indication of double counting, because the EOM for other mesons are all relate to their BSE. Only if the BSE is valid can we have an EOM.

Apart from the present way of introducing mesons into the theory, one can further consider adding in differentials of the $\Pi$ field in Eq.~(\ref{phiinsert}). These fields correspond to excited meson states and the analysis for them follows similarly as presented herein.
\section*{Acknowledgments}
This work was supported by the National Science Foundation of China (NSFC) under Grant No. 11475092.

\end{document}